# Imaging dark charge emitters in diamond via carrier-to-photon conversion

Artur Lozovoi[1], Gyorgy Vizkelethy[3], Edward Bielejec[3], and Carlos A. Meriles[1,2,†]

[1]*Dept. of Physics, CUNY-City College of New York, New York, NY 10031, USA.*
[2]*CUNY-Graduate Center, New York, NY 10016, USA.*
[3]*Sandia National Laboratories, Albuquerque, New Mexico 87185, USA.*
[†]Corresponding author. E-mail: cmeriles@ccny.cuny.edu

**ABSTRACT**: The application of color centers in wide-bandgap semiconductors to nanoscale sensing and quantum information processing largely rests on our knowledge of the surrounding crystalline lattice, often obscured by the countless classes of point defects the material can host. Here we monitor the fluorescence from a negatively charged nitrogen-vacancy (NV⁻) center in diamond as we illuminate its vicinity. Cyclic charge state conversion of neighboring point defects sensitive to the excitation beam leads to a position-dependent stream of photo-generated carriers whose capture by the probe NV⁻ leads to a fluorescence change. This "charge-to-photon" conversion scheme allows us to image other individual point defects surrounding the probe NV, including non-fluorescent "single-charge emitters" that would otherwise remain unnoticed. Given the ubiquity of color center photo-chromism, this strategy may likely find extensions to material systems other than diamond.

**KEYWORDS**: Diamond, nitrogen-vacancy centers, color centers, optical microscopy, charge dynamics, carrier capture.

## INTRODUCTION

As applications of color centers to precision sensing and quantum information processing continue to grow (*1*), the charge state of a point defect — ultimately defining its spin and optical properties — has emerged as a valuable experimental handle. A nice illustration is the case of the nitrogen vacancy (NV) center in diamond, where the interplay between spin, photon emission, and charge state has been exploited to demonstrate optical ionization conditional on its starting spin state (*2*). Since the charge state is more resilient to optical illumination than the spin, this "spin-to-charge conversion" (SCC) strategy has, in turn, been used to boost spin detection sensitivity beyond that possible via standard optical readout (*3-8*) or to demonstrate long-term spin-state storage robust to laser excitation (*9,10*).

Continuous green illumination of an NV leads to cyclic ionization and recharging, thus allowing us to think of the point defect as a local "charge pump" injecting free carriers into the conduction and valence bands as it consecutively changes its charge state from negative to neutral and vice versa (*11*). Because the number of diffusing carriers can be correlated to the defect's spin state via SCC, this process has been exploited to demonstrate electrical spin readout down to individual NVs (*12*).

Carrier recapture far from the injection site is typically marked by a drastic, long-lived change in the spin and photon emission properties of the trapping defect, meaning that, even in the absence of collecting electrodes, defect-assisted carrier generation and capture can serve as a practical tool, this time to shed light on the microscopic composition of the host crystal. Indeed, experiments articulating local laser excitation of NVs followed by non-local fluorescence imaging reveal the formation of well-defined charge state patterns whose response as a function of the excitation laser duration, intensity, and wavelength reveals valuable information on the dynamics of carrier transport and capture (*13-15*), including the formation of space charge potentials (*16*). Most recently, these ideas were extended to the limiting case where the free carrier source and/or target trap are formed by individual point defects (*17, 18*), a geometry that can potentially be exploited to establish a quantum bus between remote qubits (*19*).

Here we shift the attention back from the carrier capturing defect to the carrier source, which we indirectly image by monitoring the charge state of a probe NV center as we vary the point of laser excitation. Starting with a pair of NVs a few microns apart, we first show single-charge emitters can be revealed individually, a result we attain by working with suitably engineered "islands" formed by few (or single) NVs. We subsequently extend this approach to image more complex color center sets comprising both nitrogen- and silicon-vacancy (SiV) centers, and show that, unlike the former, the latter do not serve as efficient charge sources under optical excitation in the visible. In the process, we uncover the presence of yet-unseen charge emitting sites, which we tentatively associate with non-fluorescent vacancy-based complexes.

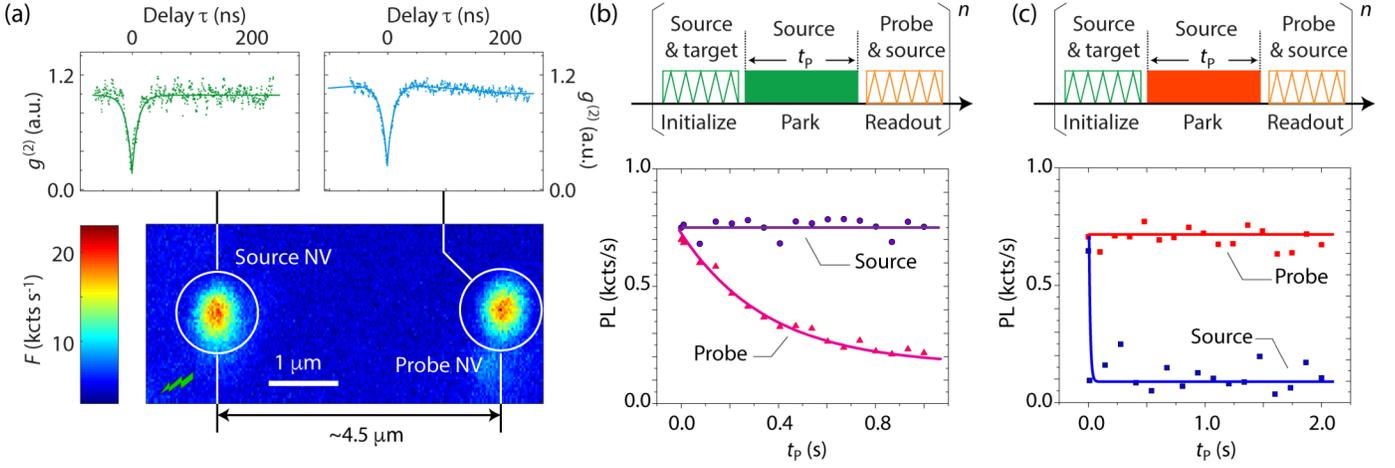

**Figure 1. Carrier transport between individual NV centers.** (a) (Main) Confocal fluorescence image under 520 nm excitation of an NV center pair approximately 4.5 µm apart. (Upper inserts) Photon autocorrelation curves for each NV; dips below 0.5 confirm the individual nature of each color center. (b) (Top) Experimental protocol. Zig-zags indicate scanning at 520 nm (green) or 594 nm (orange) while the solid rectangle indicates a 520 nm, 1 mW park at the source NV. (Bottom) Observed photoluminescence (PL) at the source and probe NVs as a function of parking time $t_P$. The probe NV fluorescence gradually decays as the integrated hole capture probability increases with $t_P$ and conversion from $NV^-$ to $NV^0$ becomes more likely. (c) Same as in (b) except that we park a 632 nm, 1.5 mW laser at the source NV (solid red block). In this case, the source NV quickly ionizes under red illumination hence preventing additional carrier generation; correspondingly, the charge state of the probe NV remains unperturbed. In (b) and (c), the green (orange) laser power during initialization (readout) is 1 mW (7 µW).

## RESULTS

### Wavelength-dependent carrier transport between individual color centers

The sample we study in the present experiments has been described in detail before (*17*). Briefly, we use a focused (~1 µm diameter) high-energy accelerator to implant N or Si ions in an electronic grade diamond crystal (DDK) with a starting nitrogen concentration of ≲ 5 ppb; the beam energy in either case (20 MeV for N and 45 MeV for Si) corresponds to a depth of ~10 µm, sufficient to rule out potential surface effects in our observations. We use variable ion beam fluences to create islands with different color center content: For nitrogen, the range goes from $1\times10^8$ ions/cm$^2$ (roughly corresponding to ~2 ions per implantation site) to $5\times10^{11}$ ions/cm$^2$, whereas for silicon we used greater fluences of up to $5\times10^{13}$ ions/cm$^2$ so as to compensate for the lower SiV conversion efficiency (*20 - 23*). Upon implantation, we followed a known annealing protocol (*24*) to simultaneously convert nitrogen and silicon atoms into NV and SiV centers. All laser excitation and confocal fluorescence measurements are carried out using a custom-made, multi-color microscope operating under ambient conditions (*10,17*).

The experiments in Fig. 1 lay out our approach: In this first example, we focus on a region of the crystal featuring a grid of single (or near-single) NVs produced by low-fluence N$^+$ implantation ($1\times10^8$ ions/cm$^2$), and choose two neighboring sites ~4.5 µm apart, each containing individual NVs, which we initialize (with 75% probability (*11*)) in the negatively charged state via green (520 nm) laser excitation. In Fig. 1b we park the green beam in one of the NVs — the free carrier "source" — for a variable time interval $t_P$ to induce multiple stochastic cycles of charge state conversion from negative to neutral and back (*11*). Each cycle leads to the injection of a conduction band electron (after NV$^-$ ionization) and a valence band hole (upon NV$^0$ recombination). Exposed to a stream of freely diffusing carriers, the negatively-charged probe NV ultimately converts to neutral as it captures a hole. Note that despite the coexistence of carriers with opposite signs, Coulombic attraction between the probe NV and the incoming hole makes the NV$^-$ to NV$^0$ transformation highly one-directional since the converse, though possible, is orders-of-magnitude less likely (*13,17*).

Since red illumination excites NV$^-$ but not NV$^0$ (whose zero-phonon line amounts to 575 nm), the dynamics at play markedly changes if we use a 632 nm laser during the park (Fig. 1c). Red light quickly ionizes the source NV — observed as an immediate fluorescence bleach — with the result that virtually no hole injection takes place; correspondingly, we see no decay in the fluorescence of the probe, whose charge state remains unchanged throughout the laser park duration.

When combined, the experiments in Fig. 1 can be seen as a change in the fluorescence of the probe contingent on the physical properties of the source, emitting carriers or not depending on the excitation wavelength. This correspondence can thus be recast as a "charge-to-photon conversion" (CPC) scheme, which, in principle, can be exploited to expose photo-activated "charge emitters". To demonstrate the idea, we return to the pair of NVs highlighted in Fig. 1 and generalize the measurement protocol to record the probe NV fluorescence for a variable position of the laser park site, a strategy we hereafter refer to as "CPC imaging".

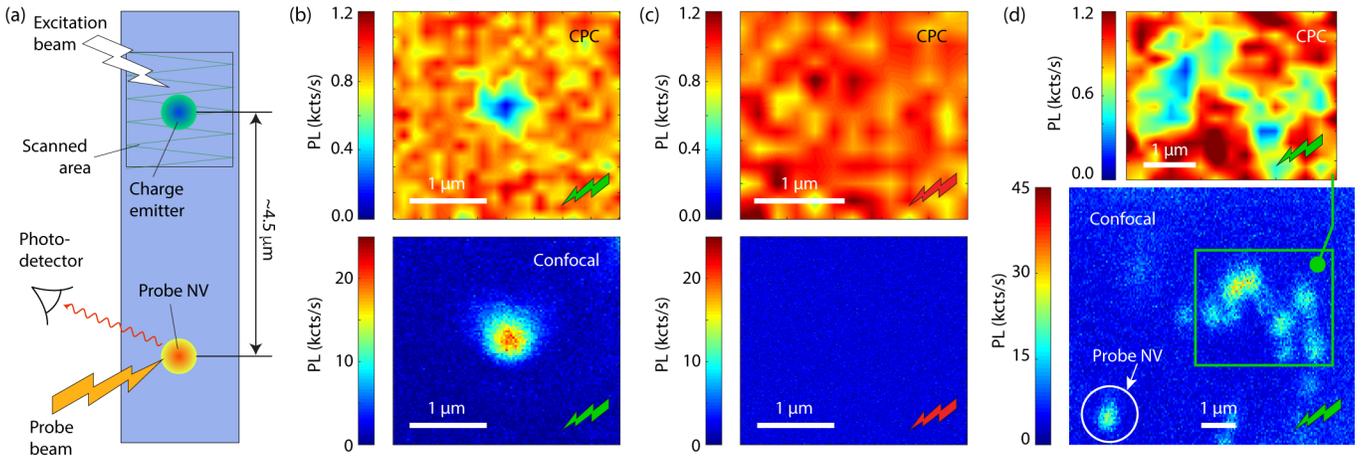

**Figure 2. Imaging an individual charge emitter via charge-to-photon conversion.** (a) Working with the NV pair of Fig. 1a, we generalize the protocols in Figs. 1b and 1c to monitor the probe NV fluorescence after a laser park at a variable position around the source NV; at each point, the park time is fixed at a value $t_P^{(0)}$ chosen to optimize the image contrast and acquisition time. (b) (Top) CPC image of the source NV using the protocol in Fig. 1b for a park duration $t_P^{(0)} = 1$ s using 520 nm, 1 mW excitation (green bolt). (Bottom) Direct confocal image of the source NV upon a green scan. (c) Same as in (b) but using a 632 nm beam throughout the CPC and confocal scans (top and bottom images, respectively); the red laser power and park duration during CPC are 1 s and 3.5 mW. (d) (Lower panel) Confocal image under 520 nm illumination of an NV "island" comprising several NVs; the white circle encloses the probe NV. (Upper panel) CPC image (green laser park, 1 mW, 3 s per point) of the area enclosed in the lower panel. We observe a good correspondence between the NV and charge emitter positions in the confocal and CPC images. The CPC pixel size is 140 nm in (b) and (d), and 200 nm in (c); CPC images have been smoothed out for clarity.

Fig. 2 summarizes the results for a fixed park duration $t_P^{(0)} = 1$ s, which we choose so as to optimize the necessary tradeoff between the probe fluorescence contrast and the image acquisition time. Since in this case the source NV is also a photon emitter, we can use standard confocal microscopy as a reference. Fig. 2b compares the results between direct fluorescence readout and CPC imaging of the same single NV using a 520 nm laser throughout the confocal scan or variable position park. We find a one-to-one correspondence, namely, the probe NV fluorescence drops at the location of the source NV but remains unchanged everywhere else. By contrast, we detect no trace of the charge emitter if we move the park laser wavelength from 520 nm to 632 nm (Fig. 2c); this observation is consistent with the rapid ionization of the source NV under prolonged red excitation, as already seen in Fig. 1.

Given the probabilistic nature of the underlying process and the relatively low hole capture cross sections of neutral or positively charged defects (*17*), CPC imaging can be extended from single charge emitters to groups containing several. This is shown in Fig. 2d where we implement our strategy in an island with multiple NVs; comparison between the CPC and confocal images — both obtained under 520 nm excitation — shows a nearly perfect correspondence between the NV locations and charge emitting sites.

**Imaging non-fluorescent charge emitters**

To benchmark our approach in yet more general scenarios, we carried out experiments in a different region of the same diamond crystal, this time in areas implanted with 45 MeV Si ions (also corresponding to a ~10 μm depth). The fluorescence image of Fig. 3a — obtained upon a 632 nm confocal scan — displays the area of interest, namely, a segment within a ~40 μm diameter circle of implanted Si ions. The presence of negatively charged SiVs is unambiguous, as demonstrated by the 737 nm zero-phonon line in the accompanying fluorescence spectrum (left of Fig. 3a).

Interestingly, 520 nm confocal microscopy also reveals a broader strip of NV centers spreading to the sides of the Si implanted area (Fig. 3b). We interpret this observation as a byproduct of Si implantation, whose relatively large size leads to multiple vacancies per ion (approximately 3 to 4 times more than nitrogen, as calculated by Stopping and Range of Ions in Matter (SRIM) simulations, not shown here for brevity). We surmise that vacancies — whose mobility is relatively high — diffuse during the annealing stage to form NV centers upon association with intrinsic substitutional nitrogen. The concentration of these "collateral" NVs — roughly 1 ppb as deduced from an average over confocal images across the implanted ring — is consistent with the above picture if we take into account the high conversion efficiency of substitutional nitrogen in bulk diamond (*25*) (of order 50%).

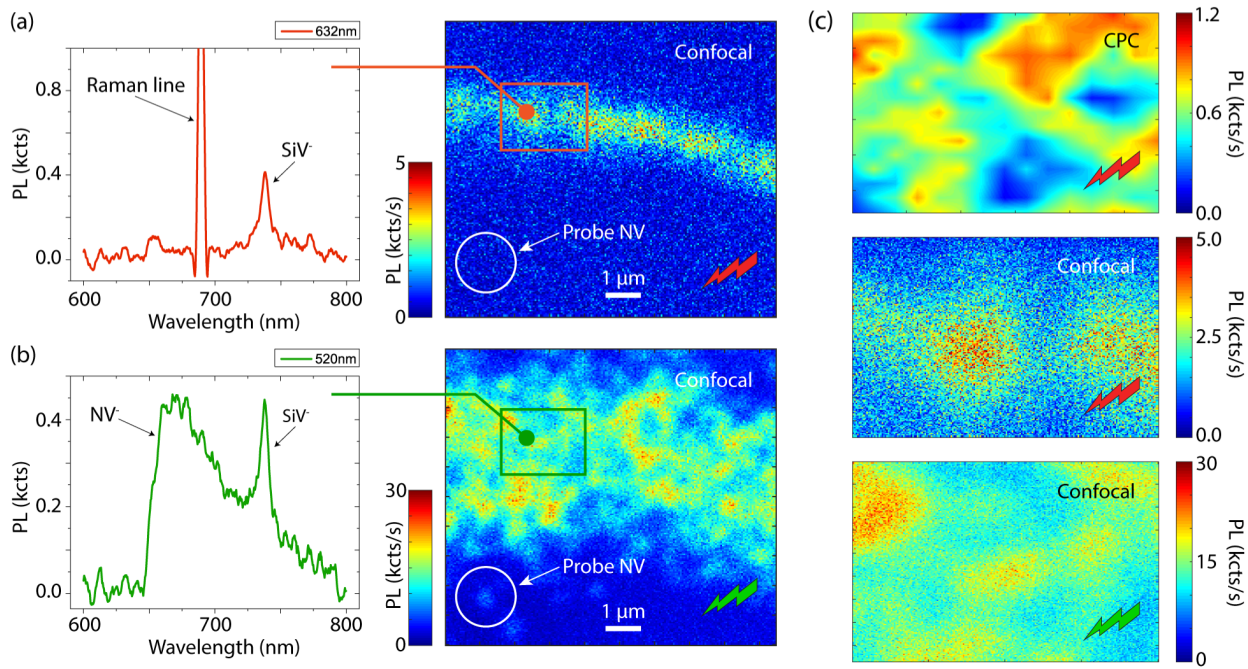

**Figure 3. CPC imaging of Si-implanted areas.** (a) (Main) Confocal image of a Si-implanted region of the diamond crystal under 632 nm excitation (Left insert) Representative fluorescence spectrum in the bright region of the image; the peak at 737 nm reveals the presence of SiV$^-$. (b) Same as above but for 520 nm excitation. "Collateral" NV$^-$ centers form around the implanted area. The white circle in the lower left corner highlights the probe NV. (c) CPC image from a red park (1 s, 4.5 mW per point) and confocal images under red and green excitation of the enclosed area in (a) and (b) (respectively, top, middle, and lower panels). We find pockets of charge emitters that do not correlate with the positions of SiV$^-$ (or NV$^-$).

Given the large fluence we chose for Si implantation ($5 \times 10^{13}$ ions/cm$^2$, orders of magnitude greater than for nitrogen, see above), it is hard to imagine NVs are the only byproduct. Fig. 3c provides initial evidence in this direction: Here we rely on a fringe NV (white circle in Figs. 3a and 3b) to CPC-image a segment of the implanted ring containing both silicon- and nitrogen-vacancy centers. We use strong 632 nm illumination during the park — which quickly leads to NV$^-$ ionization with (virtually) no hole generation — implying that NV centers in the imaged area remain invisible. We identify, nonetheless, multiple sites where the probe NV fluorescence bleaches, in the process exposing the presence of red-laser-driven pockets of charge emitters. Comparison with the confocal fluorescence image resulting from red excitation — selectively sensitive to SiV$^-$, middle panel in Fig. 3c — suggests that carriers do not stem from negatively charged SiVs. Photo-induced transformations cycling the charge state between SiV$^0$ and SiV$^-$ — presumably yielding low levels of fluorescence — are still plausible, but since we also identify this class of charge emitters in areas of the crystal presumably devoid of silicon (see below), we deem this possibility unlikely. Further, by examining a 520-nm confocal microscopy image of the same region (lower panel in Fig. 3c), we find these charge emitters are mostly located at sites with no NV fluorescence (hence allowing us to rule out unaccounted mechanisms of NV charge emission under red illumination).

These observations are revealing in several ways: Silicon-vacancy centers are known to exist in multiple charge states (*26-29*) — ranging from SiV$^{2-}$ to SiV$^+$ — but since their photochromism is poorly understood it is not a priori clear whether their charge state cycles under red excitation (particularly because red light can excite both SiV$^-$ and SiV$^0$). On the other hand, the mismatch between the CPC and confocal images — both for SiVs and NVs — hints at a distinct type of point defect, cycling its charge under red excitation without generating fluorescence. We are not aware of prior experiments able to selectively reveal similar fluorescence-free, light-driven charge emitters with diffraction limited resolution.

The results in Fig. 4 confirm and further these ideas: In this case, we move the field of view to an area close to (but away from) the Si implants (Figs. 4a and 4b). Confocal microscopy under 632 nm excitation shows that this section of the crystal is virtually devoid of fluorescence (Fig. 4d). CPC imaging, however, shows a completely different landscape, suggesting dark carrier emitters are widespread; note that, as in Fig. 3c, there is no correlation with the NV sites (see 520-nm confocal microscopy in Fig. 4e), once again ruling out this color center as the carrier-source under red light.

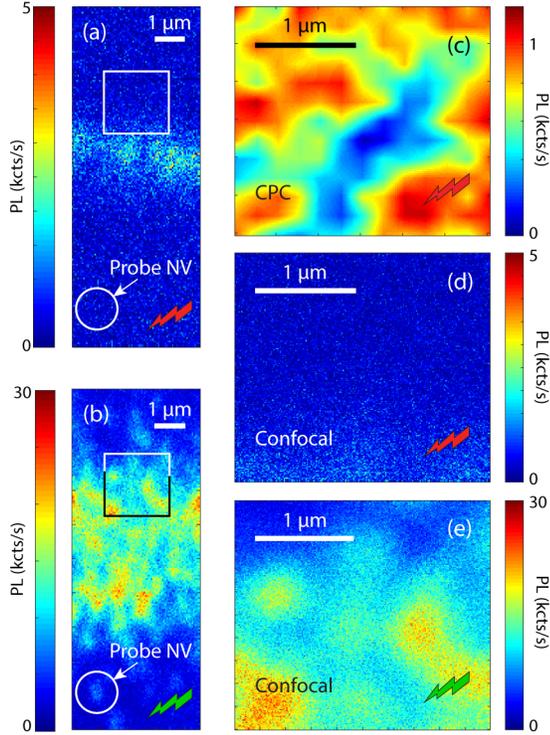

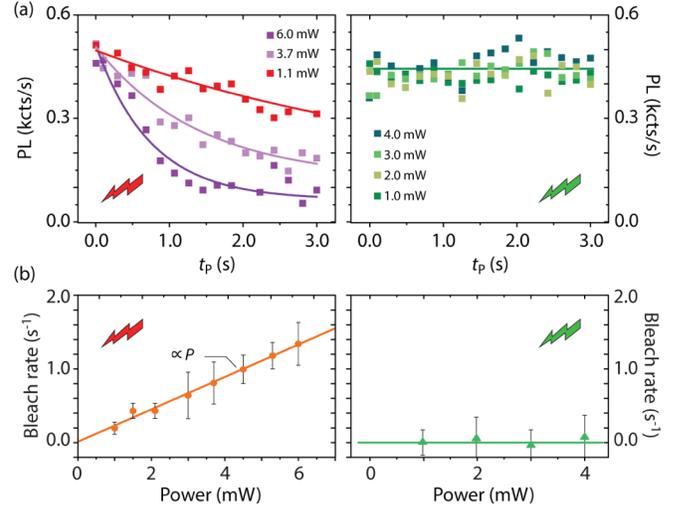

**Figure 4. Imaging dark charge emitters.** (a,b) Confocal images under red and green excitation, respectively. The white circle indicates the position of the probe NV. (c) CPC image of the enclosed area in (a) under a red laser park. (d,e) Confocal images of the same area under red and green excitation, respectively.

**Figure 5. Non-fluorescent charge emitter response at different wavelengths.** (a) Probe NV photo-luminescence as a function of the park power on a dark charge emitter 7 μm away. In these experiments, we apply the protocol in Fig. 1c (red park, left) or Fig. 1b (green park, right) for various laser powers. Solid lines are exponential fits. (b) Probe NV photo-ionization rate as a function of the applied park laser power using red or green excitation (left- and right-hand panels, respectively). Solid lines are linear fits.

Without attempting an in-depth examination of the type of emitter at play — beyond the scope of this article — CPC detection can provide some initial clues: This is shown in Fig. 5 where we park the laser beam at a site known to host a charge emitter but with no SiV or NV fluorescence, and monitor a probe NV 7 μm away as a function of time for different red laser intensities. From the linear relation between the NV fluorescence bleach rate and the applied laser power (see left-hand panels in Figs. 5a and 5b), we infer that ionization and recharging of the source charge emitter is a one-photon process, somewhat reminiscent of that inferred from observations in other chemical-vapor-deposition diamonds (*30*). Interestingly, we find almost no change in the probe NV fluorescence upon a green laser park at the same site (right-hand panels in Figs. 5a and 5b), indicating the charge state cycling stops at shorter wavelengths. Note that while less intuitive, this behavior is also seen in other emitters such as the NV (whose charge state interconversion becomes inefficient when illuminated with blue light (*11*) ).

Although the information presently at hand is insufficient to ascertain the microscopic nature of these dark emitters, the observation of collateral, Si-implantation-induced NVs suggests the concomitant formation of vacancy-related complexes, some of which we can already examine in light of the above constraints. Concrete illustrations are the V (*31*) and $N_2V$ (*32-34*) centers, the latter arguably stemming from the binding of a vacancy with a pair of adjacent substitutional nitrogens (the symbol V denotes a diamond vacancy). Both defect types are known to exhibit photochromism (*35*) between the negative and neutral charge states, but interconversion under red illumination seems unlikely due to the unfavorable energy scales at play ($V^-$ and $N_2V^-$ show zero-phonon lines at 394 and 503 nm, respectively). Given the many alternative possibilities, however, additional work will be needed to more closely examine other candidate photochromic defects — such as complexes integrating nitrogens and protons (*36-40*) — which cannot be ruled out yet.

## DISCUSSION

In summary, using an NV center as a charge probe we demonstrated the ability to image individual or small groups of NVs microns away via changes in the probe fluorescence stemming from the capture of photo-generated holes. Applying this approach to areas of the diamond crystal previously implanted with Si ions, we found that SiV centers do not undergo photochromism under red excitation. These same observations, however, revealed the presence of previously unseen pockets of non-fluorescent point defects presumably undergoing charge state cycles of ionization and recombination under red illumination (but not green). The physical nature of these charge emitters is presently unclear though the use of CPC detection already imposes some constraints on future models. In this vein, we see CPC as a technique akin to photo-luminescence excitation spectroscopy (PLE), both reporting on the integrated number of quantum particles — charge carriers or photons — produced by a source under illumination of a given color. Correspondingly, extending the above results to systematically monitor the system response as a function of the excitation wavelength would amount to spectrally fingerprinting the carrier emitter. The latter, in turn, could be exploited to separate different classes of point defects very much as in PLE.

Given the ubiquity of photochromism and carrier capture, it is reasonable to assume the present approach will also find use in the investigation of other semiconductor platforms of interest. An immediate example is the case of silicon carbide, known to host color centers whose charge properties are similar to those found in diamond (*41-43*). Another complementary direction is the combined application of sub-diffraction techniques such as STED (*44,45*) so as to ionize and image carrier capture with higher spatial resolution, down to ~10 nm. Whether or not this enhanced form of CPC can be applied to monitor photo-activated charge transport in macromolecules is an intriguing question worth investigating in further detail. Lastly, while short-range carrier capture is likely to render photo-current detection considerably less efficient (particularly if the collecting electrodes are ~10 μm apart or more), the use of external fields might prove nonetheless rewarding as a strategy to shed light on carrier propagation and/or alter its capture probability.

**MATERIALS AND METHODS**

We carry out all experiments using a multi-color confocal microscope with an oil-immersion objective (NA=1.3) adapted to focus all incoming beams and collect the sample photo-luminescence. Green (520 nm), red (632 nm), and orange (594 nm) lasers are coupled into a single-mode fiber in order to ensure a similar beam intensity profile and best possible overlap of the excitation volumes. A combination of long-pass, short-pass, and band-pass optical filters is used when separating the photo-luminescence from $SiV^-$ and $NV^-$. We readout the NV charge state by scanning an area around the probe NV with a low-power (7 μW) 594nm orange excitation beam, and then integrate the photo-luminescence from the pixels corresponding to the position of the probe NV. By slightly increasing the integrated area of the scan, we can account for potential instabilities in the sample position, which otherwise would have compromised the contrast in the single-point-illumination readout.

At high red laser power and/or short distances between the target area and the probe NV, it is possible to bleach the probe $NV^-$ photo-luminescence due to ionization caused by the red beam tail and by scattered red light. This hinders the bleaching caused by hole capture and is therefore unwanted in the CPC protocol. We circumvent this effect by recording reference bleaching curves where the target is at the same distance from the probe NV but in a section of the crystal with no charge emitters. CPC-compatible conditions are those where no bleaching is observed (the case in all our experiments).

**Acknowledgements:** We acknowledge fruitful discussion with Yun-Heng Chen and Marcus Doherty. **Funding.** A.L. and C.A.M acknowledge support from the National Science Foundation through grant NSF-1914945. The authors acknowledge the facilities and research infrastructure support of the NSF CREST-IDEALS, NSF grant number HRD-1547830. Ion implantation work to generate the NV and SiV centers was performed, in part, at the Center for Integrated Nanotechnologies, an Office of Science User Facility operated for the U.S. Department of Energy (DOE) Office of Science. Sandia National Laboratories is a multi-mission laboratory managed and operated by National Technology & Engineering Solutions of Sandia, LLC, a wholly owned subsidiary of Honeywell International Inc., for the U.S. Department of Energy's National Nuclear Security Administration under contract DE-NA0003525.This paper describes objective technical results and analysis. Any subjective views or opinions that might be expressed in the paper do not necessarily represent the views of the DOE or the United States Government. **Author contributions:** A.L. and C.A.M. conceived the experiment. G.V. and E.B. carried out the ion implantation and A.L. conducted the optical microscopy experiments. C.A.M. supervised the overall research effort and wrote the manuscript with input from all authors. **Competing interests:** All authors declare that they have no competing interests. **Data and materials availability:** All data needed to evaluate the conclusions in the paper are present in the article. Additional data related to this paper may be requested from the authors. All correspondence and request for materials should be addressed to C.A.M. (cmeriles@ccny.cuny.edu).